\documentclass[twocolumn]{aastex631}

\usepackage[utf8]{inputenc}
\usepackage[T1]{fontenc}

\shortauthors{Vipin Sudevan \& Pisin Chen}

\usepackage{apjfonts}
\usepackage[symbols,nogroupskip,nonumberlist]{glossaries-extra}

\shorttitle{${\tt PUREPath-B}$}

\begin{document}

\title{PUREPath-B: A Tessellated Bayesian Model for Recovering CMB B-modes over Large Angular Scales of the Sky}
\author{Vipin Sudevan}
\affiliation{Leung Center for Cosmology and Particle Astrophysics, National Taiwan University, Taipei 10617, Taiwan}

\author{Pisin CHen}
\affiliation{Leung Center for Cosmology and Particle Astrophysics, National Taiwan University, Taipei 10617, Taiwan}
\affiliation{Department of Physics and Center for Theoretical Sciences, National Taiwan University, Taipei 10617, Taiwan}
\affiliation{Graduate Institute of Astrophysics, National Taiwan University, Taipei 10617, Taiwan}
\affiliation{Kavli Institute for Particle Astrophysics and Cosmology, SLAC National Accelerator Laboratory, Stanford University, Stanford, California 94305, USA}

\begin{abstract}
{We introduce a comprehensive, custom-developed neural network, the ${\tt PUREPath-B}$, that yields a 
posterior predictive distribution of Cosmic Microwave Background (CMB) B-mode signal conditioned on the 
foreground contaminated CMB data and informed by the training dataset. Our network  employs nested   
probabilistic multi-modal U-Net framework, enhanced with probabilistic ResNets 
at skip connections and seamlessly integrates    
Bayesian statistics and variational  methods to minimize the foreground and noise contaminations.  
During training, the initial prior distribution over network parameters evolves into approximate posterior 
distributions through Bayesian inference,  constrained by the training data.
From the approximate joint full posterior of the model parameters, our network infers a predicitve CMB posterior 
during inference and yields summary statistics such as 
predictive mean, variance of the cleaned map. 
The predictive standard deviation provides an interpretable measure of per-pixel uncertainty in 
the predicted mean CMB map. For loss function, we use a linear combination of KL-Divergence loss and weighted MAE—which 
ensures that maps with higher amplitudes do not dominate the loss disproportionately. 
Furthermore, the results from the cosmological parameter estimation 
using the cleaned B-mode power spectrum, along with its error 
estimates demonstrates our network minimizes the foreground  
contaminations effectively, enabling 
accurate recovery of tensor-to-scalar ratio and lensing amplitude.}
\end{abstract}


\section{Introduction}
\label{into}

The CMB serves as a cornerstone for understanding the origin and 
evolution  of the universe, offering precise insights into the initial conditions, composition, 
and evolution of the universe~\citep{Planck:2018vyg}.  
Among its polarization components, the B-modes, which are exclusively sourced by 
tensor perturbations~\citep{Seljak:1996gy,Kamionkowski:1996zd,Kamionkowski:1996ks}, offers a 
unique probe for the primordial 
gravitational waves, the amplified quantum fluctuations of the gravitational field~\citep{FABBRI1983445,MUKHANOV1992203,Grishchuk_1977}
during the inflationary epoch~\citep{PhysRevD.23.347,Starobinsky:1979ty}. 
The amplitude of B-modes, quantified by 
the tensor-to-scalar ratio $r$, encodes information about the energy scale of inflation, 
potentially reaching energy levels inaccessible 
to modern particle accelerators. Therefore, detecting B-modes provides a unique opportunity to test the 
inflationary paradigm and explore quantum gravitational effects at unprecedented energy scales.

However, accurately estimating B-modes is a significant challenge due to their faintness as 
compared to astrophysical foreground emissions, 
instrumental noise, etc. Polarized emissions from galactic synchrotron radiation, thermal dust, etc., 
as well as lensing-induced B-modes 
from large-scale structures, and detector noise dominates the observed signal.  
A robust detection of  
B-mode signal requires  advanced component separation techniques, extreme instrument sensitivity,  
control over systematic errors  and precise and 
sensitive measurements over a wide range of angular scales and frequencies. Ongoing and 
future CMB experiments~\citep{CORE:2017ywq,LiteBIRD:2022cnt,Adak:2021lbu,Li:2017drr,Hui:2018cvg,CMB-S4:2022ght}, 
aim to achieve the sensitivity and frequecny range required to isolate primordial B-modes from contaminants.

Recent decades, several foreground minimization and component separation methods~\citep{Eriksen2008, Eriksen2008a, Land:2005cg,Jaffe:2006fh,2003ApJS..148...97B,2004ApJ...612..633E,Tegmark2003,2009A&A...493..835D,sudevan2017improved,Sudevan:2018qyj,Sudevan:2017una,Taylor:2006otn,2013A&A...558A.118H} are developed to minimize the effects of foreground contaimnations and thereby 
providing cleaned CMB maps.  With the remarkable achievements 
in the field of machine learning (ML)~\citep{5392560} in tackling 
various complex real-world challenges, ML based techniques are investigated to minimize the foregrounds 
in  CMB data~\citep{Petroff:2020fbf,Wang:2022ybb,Casas:2022teu,Yan:2023bjq, 2025arXiv250209071Y, Pal:2024cir, Sudevan:2024hwq, defferrard2020deepsphere}.  Some of the other interesting applications of ML in CMB are~\cite{Farsian:2020adf,Pal:2022woh,Pal:2024cir, Adams:2023uod,Petroff:2020fbf,Lambaga:2025nbw,Melsen:2024eiw}
and others. 
In conventional ML framework,  a model is trained to understand the relationship  
between inputs and outputs by fixing the model parameters 
(${\omega}$), the weights and biases. 
Probabilistic neural networks~\citep{SPECHT1990109} like 
Bayesian neural  networks are developed to address a crucial shortcoming 
of the standard deep neural networks 
by incorporating uncertainty considerations 
in both model and data.

We present a Bayesian network, ${\tt PUREPath-B}$, which utilizes probabilistic layers in order to 
capture the uncertainty over weights by incorporating prior knowledge of the parameters and 
propagating it through the network to model the prediction uncertainty. 
Our network, designed to minimize foregrounds in the simulated CMB {\bf B}-mode maps,  
is best characterised as a tesellated {\bf P}robabilistic, nested framework composed 
of $n_{\tt reg}$ (total numebr of regions) probabilistic {\bf U}-Nets~\citep{2015arXiv150504597R}. Each U-Net 
consists of $n_{\tt map}$ (total number of maps at different frequencies) encoders to facilitate a multi-modal learning,  
a latent space, and a decoder with probabilistic {\bf R}esNets~\citep{2015arXiv151203385H}
 in the {\bf E}xplanding 
{\bf path}way of each decoders at skip connections and an output defined by a diagonal multivariate normal distribution.

\section{Formalism}
\label{formalism}

In a Bayesian network, the model parameters (${\omega}$)   are   
stochastic and are sampled from respective posterior distributions 
$P(\omega | \mathcal{D})$, where,  $\mathcal{D}$ is the 
training dataset 
with $N$ simulated sets of foreground contaminated CMB maps ${\bf X}_j = \{{\bf x}_{i,\,j}\}_{i=1}^{n}$ at $n$ frequencies as inputs, 
and $N$ output ${\bf y}_j$ (the simulated CMB map for each set $j$) i.e., 
$\mathcal{D} = \Bigl\{\{{\bf x}_{i,\,j}\}_{i=1}^{n}, {\bf y}_j \Bigl\}_{j=1}^{N}$.
For a given prior $P({\omega})$ and $\mathcal{D}$, the posterior $P(\omega | \mathcal{D})$~\citep{graves2011practical} 
is updated using the Bayes' principle as follows:  
\begin{equation}
P(\omega|\mathcal{D}) = \frac{P(\mathcal{D}|\omega)P({\omega})}{P(\mathcal{D})} \, .
\label{bayes}
\end{equation}
$P(\mathcal{D}|\omega) = \prod_{i=1}^{N} P({\bf y}_i | {\bf X}_i, \omega)$ is the  
$likelihood$ of the observing ${\bf y}_i$ for a 
given ${\bf X}_i$ and $\omega$. 
Estimating exact $P(\omega| \mathcal{D})$ using Bayesian inference~\citep{Tipping2004} 
is not practical due to the integral over all feasible $\omega$, hence  
approximate methods like Monte Carlo, Variational Inference (VI)~\citep{2013arXiv1312.6114K, JMLR:v14:hoffman13a, 2015arXiv150901631K} 
are  used. VI approximates exact distribution with a tractable 
surrogate or varaitional distribution $Q_\theta({\omega})$ with 
parameters $\theta$,  
 optimized to 
minimize the  Kullback-Leibler (KL) 
divergence $\mathbb{D}_{KL} (Q_\theta({\omega}) || P({\omega} | \mathcal{D}))$~\citep{kullback1951information}, 
\begin{align}
\mathbb{D}_{KL} (Q_\theta({\omega}) || P({\omega} | \mathcal{D})) = & \mathbb{E}_Q [\log Q_\theta({\omega})] - \mathbb{E}_Q [\log P({\omega} | \mathcal{D})] \nonumber \\ & + \log(P(\mathcal{D})) \, .
\label{kl}
\end{align}
In this work,  we minimize 
\begin{equation}
\mathbb{L} = \frac{1}{N} \sum_{i}^{N}w_i |{\bf y}_i^{true} - {\bf {\bar y}}_i^{pred}| + \beta \mathbb{D}_{KL} (Q_\theta(\omega) || P(\omega | {\mathcal D})) \, ,
\label{loss}
\end{equation}
where, $w_i = \frac{1}{A_i + \epsilon} \Big/ \frac{1}{N} \sum_{j}^{N} \frac{1}{A_j + \epsilon}$ and 
$A_i = \sqrt{\frac{1}{N_{\tt pix}}\sum_{j=1}^{N_{\tt pix}}\big({\bf y}_{ij}^{true}\big)^2}$. 
$\epsilon$ is a small constant ($10^{-6}$) and ${\bf {\bar y}}_i^{true}$ is the target CMB map with 
$N_{\tt pix}$  number of pixels while   
$\beta$ is a tunable hyper parameter to weight the KL-Diveregence. The weighted MAE prevents the loss 
from being dominated by high-amplitude maps, since we use B-mode maps with $r$-value ranging from $10^{-4}$ to $10^{-2}$. 
During training, the loss function, $\mathbb{L}$, is 
minimized to find the optimal values of $\theta$ of the weight distributions. The optimization relies 
on the Flipout technique~\citep{2018arXiv180304386W}, which provides efficient and low-variance weight 
perturbations, ensuring stable and effective training. 

\begin{figure}
 \centering
\includegraphics[scale=0.17]{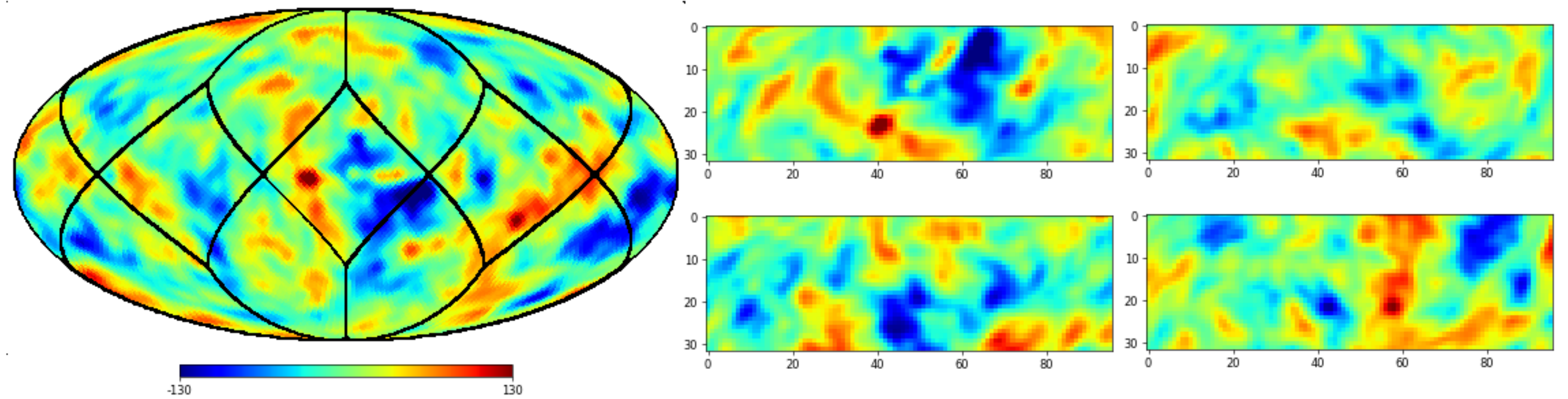}
\caption{A simulated CMB map is displayed in the left panel in HealPix Mollewide 
projection. The map is divided with solid black lines into 12 equal area regions with dimension 
$N_{\tt side}\, \times\, N_{\tt side}$. Each low-latitude region is stiched together with 2 neighboring 
higher-latitude areas to form  4 independent plane images of dimension $N_{\tt side}\, \times\, 3N_{\tt side}$ 
displayed in the middle and right panel.    }   
\label{inp_map}
\end{figure}

Once trained, the predictive CMB posterior ${\hat P}({\bf y}^\ast|{\bf X}^\ast, \mathcal{D})$ 
of the predicted cleaned CMB map (${\bf y}^\ast$) 
conditioned on ${\bf X}^\ast$ and $\mathcal{D}$, 
is estimated using the variational posterior 
$Q_\theta({\omega|\theta^\ast})$ 
of the model parameters as follows, 
\begin{align}
{\hat P}({\bf y}^\ast|{\bf X}^\ast, \mathcal{D}) 
&\approx \int {\hat P}({\bf y}^\ast|{\bf X}^\ast, \omega) \mathcal{Q}_\theta(\omega|\theta^\ast) d\omega \, .
\label{samp}
\end{align}
This integral can be performed numerically, 
\begin{equation}
{\hat P}({\bf y}^\ast |{\bf X}^\ast, \mathcal{D}) \approx \frac{1}{S} \sum_{s=1}^{S} {\hat P}({\bf y}^\ast | {\bf X}^\ast, \omega^s)  \, , 
\label{posterior}
\end{equation}
where, ${\hat P}({\bf y}^\ast | {\bf X}^\ast, \omega^s) = \mathcal{N}({\bf y}^\ast | \mu_{\tt CMB}({\bf x}^\ast, \omega^s), \sigma_{\tt CMB}({\bf x}^\ast, \omega^s))$~\citep{abdar2021review} is a predictive distribution 
corresponding to $\omega^s$ sampled from  $Q_\theta({\omega|\theta^\ast})$  i.e., $\omega^s\,\sim\,Q_\theta({\omega|\theta^\ast}$).   
$s$ and $S$ represents the index of the sampling 
step and the total number of MC samplings respectively, while,  
$\mu_{\tt CMB}$ and $\sigma_{\tt CMB}$ are 
the sample mean and standard deviation estimated from the samples generated in  
the step $s$. 
Using ${\hat P}({\bf y}^\ast |{\bf x}^\ast, \mathcal{D})$, 
we compute the summary statistics such as the predictive mean, variance 
and standard deviation as follows:
\begin{align}
\mathbb{E}[{\bf y}^\ast | {\bf x}^\ast, {\mathcal D}] &\approx \frac{1}{S} \sum_{s=1}^{S} \mu_{\tt CMB}^s \label{pmean}\, ,\\
Var[{\bf y}^\ast |{\bf x}^\ast, \mathcal{D}]  &\approx \frac{1}{S} \sum_{s=1}^{S}\Bigl((\mu_{CMB}^s - \mathbb{E}[{\bf y}^*])^2 + (\sigma_{\tt CMB}^s)^2 \Bigl)\label{pvar}\, , \\
\sigma[{\bf y}^\ast |{\bf x}^\ast, \mathcal{D}] &= \sqrt{Var[{\bf y}^\ast |{\bf x}^\ast, \mathcal{D}]} \label{pstd}\, .
\end{align}


\section{Network Architecture}
\label{Architecture}

\begin{figure}
\includegraphics[scale=0.33]{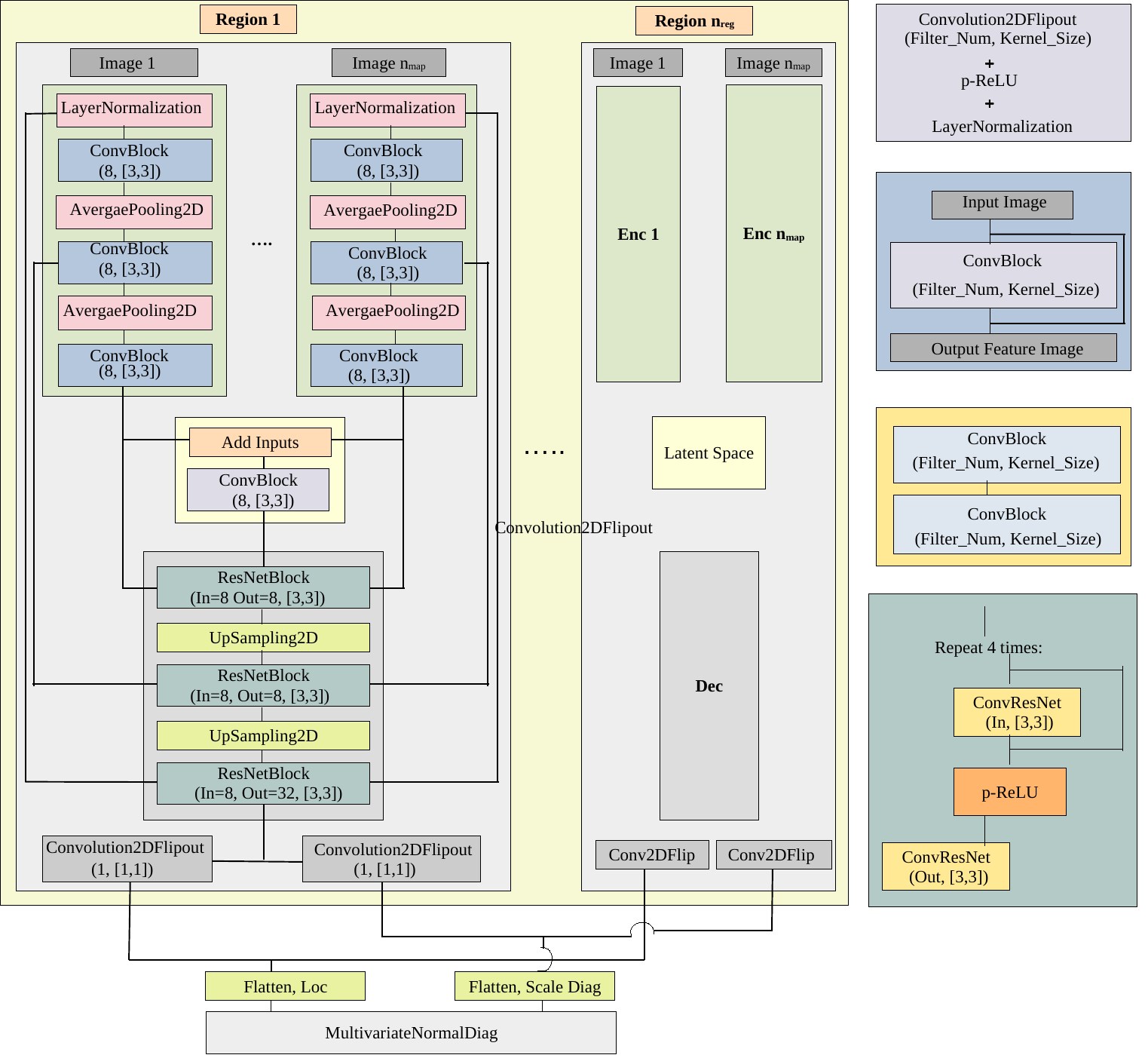}
\label{net}
\caption{A pictorial representation of our ${\tt PUREPath-B}$ Model. Input maps corresponding to each region 
is processed by its own multi-encoder-latent-decoder
units. Each encoder block in turn has separate encoder units to process each input map 
 in that region merging into a single 
latent space  and a decoder unit. All the feature maps from the decoders forms the input to  
a MultivariateNormalDiag layer, which is the output layer of our model. }
\label{net_pic}
\end{figure}

The ${\tt PUREPath-B}$ architecture, displayed in  Figure~\ref{net_pic}, is a nested combintaion of 
multiple U-Nets to handle images from multiple regions in the sky. Each U-Net in turn comprises of as many 
encoders as the numebr of input frequency maps. Our network consists of    
Tensorflow Probability~\citep{tensorflow2015-whitepaper}  
Convolution2DFlipout layers with kernel size $3\times3$ and  
strides set as 1 as the feature extracting layers, followed by 
{\it parametric}-RELU ({\it p}-RELU) activation and LayerNormalization.
The Convolution2DFlipout layers are initialized with hierarchical priors that 
specify a multivariate normal distribution for the layer's weights, allowing the mean and variance 
to be learned from the data along with mean-field normal posteriors, both provided by  
Tensorflow. At each encoder layer, the generated feature maps are first concatenated 
with input maps to that layer  and then downsampled using Tensorflow AvergaePooling2D.  
The feature maps from final layer of all encoders in a multi-modal U-Net are combined 
to from a high-dimensional, abstract representation of inputs images in  
latent space. These are passed to decoder along with the 
encoder outputs to form skip-connections. In the decoder, 
each layer's input consists of upsampled output from the previous decoder 
layer, the corresponding encoder output at same resolution. 
Each  skip connection incorporates a ResNet Block, which consists of two  convolution 
operations each 
followed by p-RELU activation and LayerNormalization. A 
residual connection is formed by adding back the input and this process is repeated 3 times. 
At the final layer, the decoder outputs are combined and flattened to parameterize a 
MultivariateNormalDiag' layer, defining a probabilistic output distribution.  
The mean represents the network prediction, and the standard deviation quantifies the 
aleotoric uncertainty.

\begin{figure*}
\centering
\includegraphics[scale=0.475]{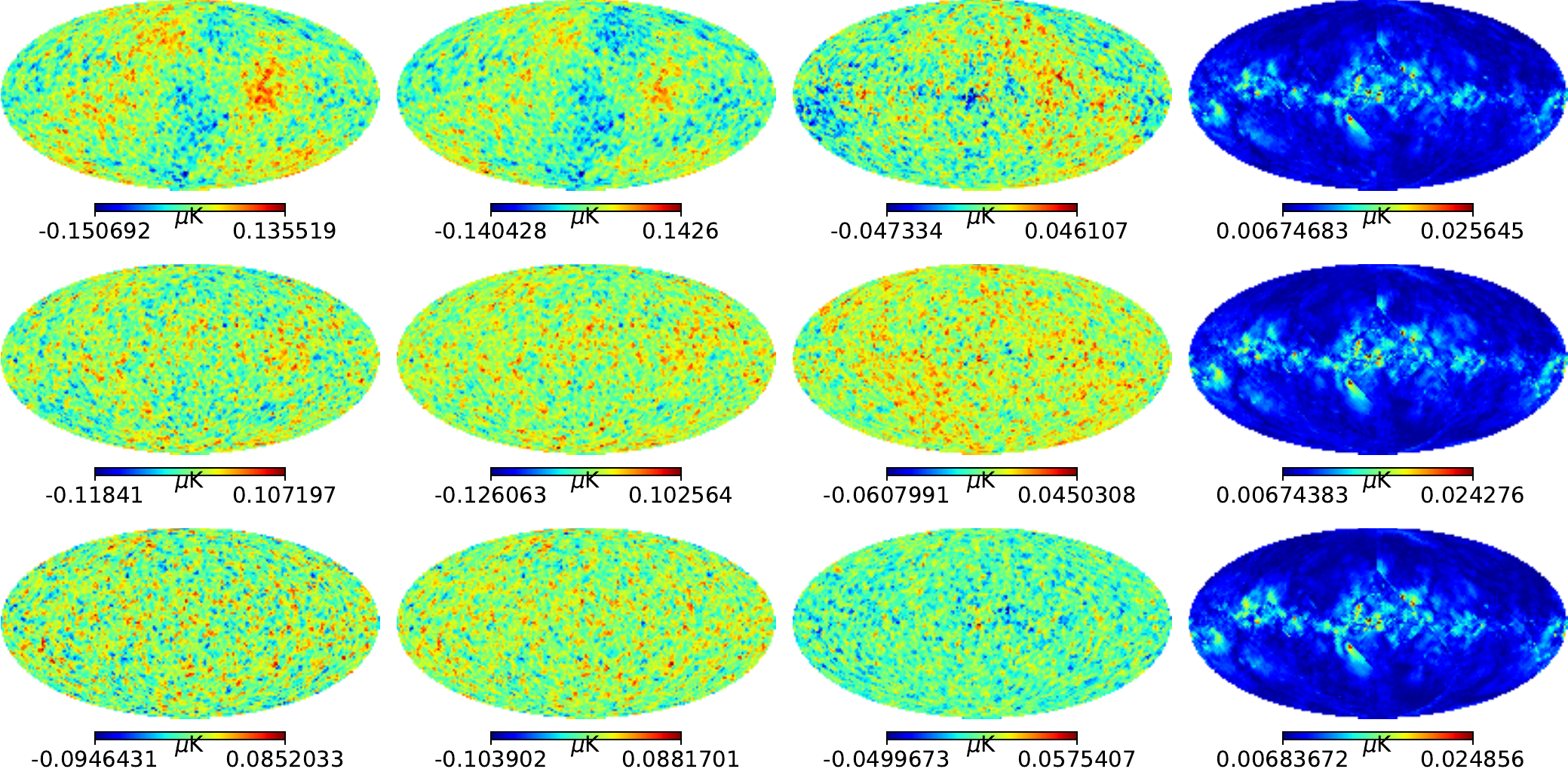}
\caption{Summary of the performance of our network in  minimizing foregrounds at three 
different r-values. In each row, the left-most panel displays the target CMB map, while the 
adjacent left-middle panels shows the ${\tt Mean\_CMap}$ predicted by our network  
after minimizing the foregrounds present in the input maps at HealpPix 
$N_{\tt side}$ = 32 and with a $1^\circ$ beam smoothing.  
The difference between our ${\tt Mean\_CMap}$ and the target CMB maps are shown in 
the right-middle panels. In the right-most panels, we show the error maps which combines both 
aleotoric and epistemic uncertainities. The top row corresonds to the results  
for r-value = $1.21\times10^{-2}$, while middle and bottom row 
corresponds to results from r-values $4.45\times10^{-3}$ and $6.45\times10^{-4}$, respectively. }
\label{simmaps}
\end{figure*}

\section{Methodology}
\label{methododlogy}
We train our network using foreground-contaminated CMB maps simulated at several 
LiteBIRD~\citep{LiteBIRD:2022cnt} frequency channels using software packages 
CAMB~\citep{2011ascl.soft02026L}, 
HealPix~\citep{Gorski:2004by}, and PySM~\cite{Thorne:2016ifb}. 
The full-sky CMB simulations follow  the $\Lambda$-cold 
dark matter framework by sampling the tensor-to-scale ratio $r$ unformily between 
$\log_{10}(0.0001)$ and $\log_{10}(0.05)$ and the Lensing Amplitude $A_L$ from a Gaussian
distribution with mean value set at 1.0 and 1$\%$ stanadard deviation 
while   
other cosmological parameters are fixed with values provided by~\cite{Planck:2018vyg}. 
Employing CAMB, we generate 3000 lensed CMB B-mode power spectra and the full-sky map using  PySM.

Foreground emission maps are simulated using PySM at the eight LiteBIRD frequencies—40, 78, 100, 140, 195, 
280, 337 and 402 GHz. For each simulation, we randomly select one model for thermal dust from the set 
$\{d0,\, d4\}$, similarly one for synchrotron emission from $\{s1,\, s2\}$, and one for 
AME from $\{a1,\, a2\}$. Consequently, the 3000 foreground simulations are generated from random 8 unique combinations 
of these foreground models. This ensures our network is trained on a wide array of complex and 
realistic foreground scenarios, closely mirroring the intricate true sky distribution.
Additionally, we generate 3000 noise realizations consistent with the noise model for 
LiteBird detectors~\citep{LiteBIRD:2022cnt}.
All the maps are generated  at HealPix $N_{\tt side} = 32$ 
with a Gaussian smoothing of FWHM $1.83^\circ$. 

To ensure seamless integration of Convolution2DFlipout layers, we transform the spherical full-sky maps to 
approximate plane images~\citep{Wang:2022ybb} by 
reordering all the maps from native HealPix Ring to Nested pixellation. We divide the resulting 
Nested maps to 12 equal area regions with dimension ($N_{\tt side} \times N_{\tt side}$) 
as shown in the left panel of Figure~\ref{inp_map} during preprocessing. 
We take one low-latitude region and combine it with 2 neighboring higher-latitude 
areas to form  4 independent 
($N_{\tt side} \times 3N_{\tt side}$) planar maps~\citep{Sudevan:2024hwq}. 
This procedure yields 4 sets of 3000 planar images for  
each frequency channel. The target full-sky CMB maps are normalized by subtracting the overall mean and 
dividing by overall standard deviation.  
We use 2900 samples for training, remaining 100 for testing, and 10$\%$ of the training 
dataset for validation. 

Training utilizes a random 80$\%$ of the training data in batches of 16 samples every epoch,  
vastly reducing the memory consumption and improving generalization.  
The Adam optimizer~\citep{2014arXiv1412.6980K} is initialized 
with learning rate 0.01 which is gradually reduced by 25$\%$ 
if validation loss stagnates for 100 epochs. 
We use Gradient tape for training with gradient accumulation every 5 steps and the 
network stops training if the validation 
loss fails to improve over a consecutive 300 epochs. 
The model weights (mean and standard deviation) corresponding to the lowest validation loss
are saved.

\section{Results}
\label{results}

\begin{figure}
 \centering
\includegraphics[scale=0.7]{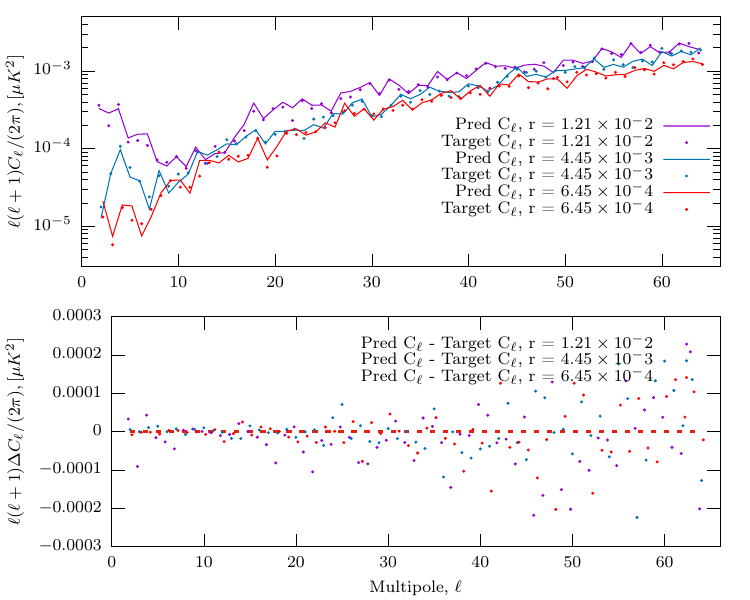}
\caption{The angular power spectrum estimated from the 
${\tt Mean\_CMap}$ for r-values $1.21\times10^{-2}$, $4.45\times10^{-3}$ and $6.45\times10^{-4}$ 
are shown in violet, blue and red lines respectively. The corresponding target map power spectra is 
shown in violet, blue and red points. We see excellent agreement between the predicted and 
target power spectra for all ranges of $r$ at all multipoles. The difference between input and 
predicted power spectra is shown in the bottom panel. }
\label{simcl}
\end{figure}
\begin{figure}
 \centering
\includegraphics[scale=0.24]{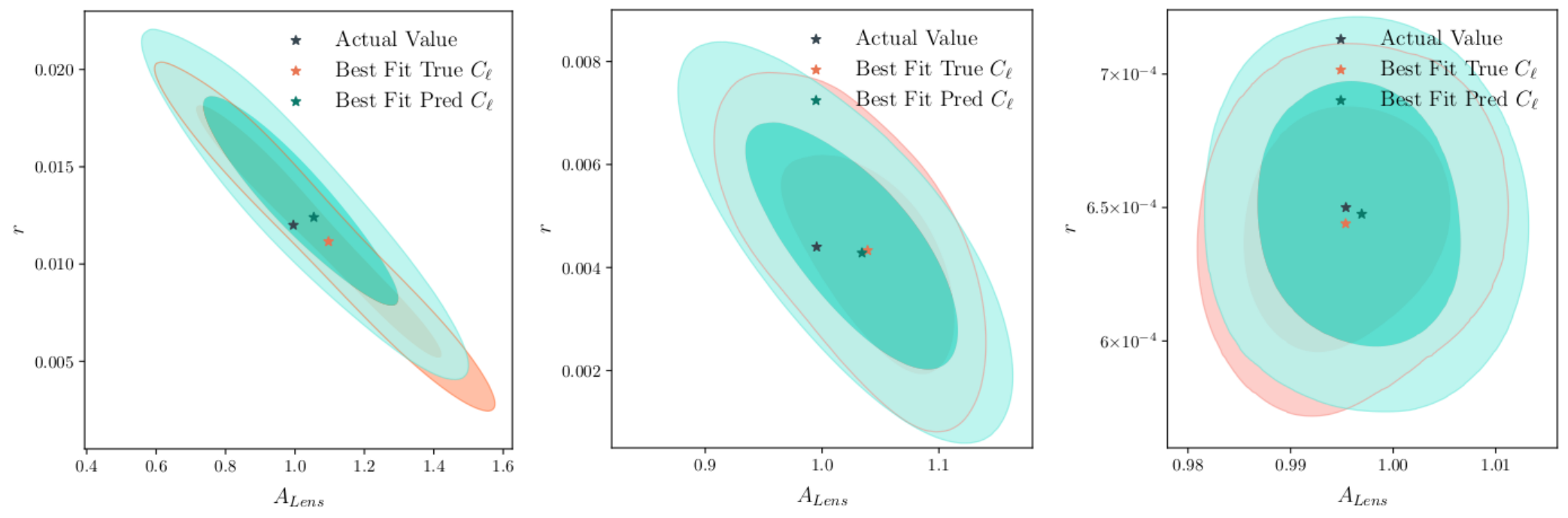}
\caption{The likelihood estimated using the map power spectrum after taking into account of the network 
uncertainty is shown in Fig.~\ref{simcl} for r-values $1.21\times10^{-2}$, 4.45$\times10^{-3}$ and $6.45\times10^{-4}$ 
are displayed in the left, middle and right panels respectively in blue contours. The red contours corresponds 
to the likelihood estimated using the target map power spectra without including network error. We see that 
the best-fit value is close to the true value shown as black point. }
\label{like}
\end{figure}
During inference, for each set of input foreground-contaminated CMB B-mode maps, we repeatedly sample 
our ${\tt PUREPath-B}$’s weights to obtain an output distribution for the cleaned CMB B‑mode map. For 
each weight sample, we extract the sample mean and standard deviation from the corresponding distribution. The 
final cleaned map (${\tt Mean\_CMap}$) is the ensemble mean of these sampled means, while the overall 
error estimate combines the network’s predicted uncertainties (the standard deviation maps from each weight
samples) with the dispersion across the sampled means, 
as described in  Eqn.~\ref{pstd}. This captures both epistemic and aleatoric contributions. 
We show results from inference for some representative random $r$ values ($r$ =  0.012, 0.0045 0.00065) 
in Figure~\ref{simmaps}, arranged from top to bottom 
in decreasing order of $r$.  In the  
left-most panel we show the target CMB B-mode map and the corresponding ${\tt Mean\_CMap}$ predicted by our network
in the middle left panel. 
In the middle right panel we show the difference between the predicted ${\tt Mean\_CMap}$s
and the input CMB maps. From the difference maps we see that out network 
predictions  agrees well with the input CMB B-mode maps. In right-most panel we show the error map which 
captures both epistemic and aleatoric contributions.
 
The full-sky power spectra of the ${\tt Mean\_CMap}$s and the input CMB B-mode maps are estimated 
after properly accounting  for the  beam and pixel window functions. 
However, we observe some residual noise bias in our estimated power spectra $C_\ell^{\tt pred}$. 
To correct this bias, we use 80 smaples from the testing data and compute for each multipole $\ell$ 
the ratio $R(\ell)=C_\ell^{\tt true} / C_\ell^{\tt pred}$. For each sample, we fit a linear model
$R(\ell) \, \sim \, m(\ell)A\, +\, c(\ell) $ where $A$ is the mean of $C_\ell^{\tt pred}$  
from $\ell\,=\,2$ to 10. Then for any new predicted spectrum with amplitude $A^{\prime}$, 
the corrected power spectrum is obtained as 
\begin{equation}
C_\ell^{\tt correct} = [m(\ell)A^{\prime} + c(\ell)]\, C_\ell^{\tt pred} \, .
\end{equation}
Since this correction is multiplicative, we propagate the uncertainty by applying the same factor to 
the error estimates (i.e., $\sigma_{C_\ell}^{\tt correct}\,=\,[m(\ell)A^{\prime}\,+\,c(\ell)]\,\sigma_{C_\ell}^{\tt pred}$). The corrected power spectra  shown in Figure~\ref{simcl} with solid lines 
representing the power spectra corresponding to ${\tt Mean\_CMap}$s shown in Figure~\ref{simmaps} with 
violet, green and red colors  for $r$ values 0.012, 0.0044, 0.00064 respectively, alongside  
the corresponding input CMB B-mode maps power spectra in points with matching colors. 
We see that estimated power spectra matches well with the input CMB power spectra at all mutlipoles, $\ell$ 
and do not show any signs of bias due to residual foreground or noise contributions. We show the 
difference between these power spectra in the bottom panel of Figure~\ref{simcl}. 

We use corrected CMB B-mode map power spectrum to constrain the primordial $r$-value and lensing 
amplitude $A_{Lens}$ through cosmological 
parameter estimation implemented with Cobaya. We use a likelihood function, $L (r, A)$ 
which is computed as the product of contributions from individual multipoles: 
\begin{equation}
L(r, A) = \prod_{\ell = \ell_{\text{min}}}^{\ell_{\text{max}}} L_\ell(r, A)\, .
\end{equation}
$L(r, A)$ accounts for uncertainty due to cosmic variance and error estimated using our network ($\sigma_{\text{Net},\ell}^2$). 
To do this we divided $L(r, A)$ into distinct regimes for low $\ell$ ($\ell$<30) and high $\ell$ as follows:

$L_\ell(r, A) =$
{\small
\begin{align}
\begin{cases}
\displaystyle
\begin{cases}
\displaystyle \int_{C_{\min}}^{C_{\max}} dC\, \frac{C^{\alpha_\ell-1}e^{-C/\theta_\ell}}{\Gamma(\alpha_\ell)\,\theta_\ell^{\alpha_\ell}} \cdot \frac{\exp\!\Bigl[-\frac{\bigl(C_\ell^{\mathrm{pred}}-C\bigr)^2}{2\sigma_{\mathrm{Net},\ell}^2+\epsilon}\Bigr]}{\sqrt{2\pi\,\bigl(\sigma_{\mathrm{Net},\ell}^2+\epsilon\bigr)}}, & \text{if } \sigma_{\mathrm{Net},\ell} \geq \text{T}, \\[2ex]
\displaystyle \frac{C_\ell^{\mathrm{pred}\,\alpha_\ell-1}e^{-C_\ell^{\mathrm{pred}}/\theta_\ell}}{\Gamma(\alpha_\ell)\,\theta_\ell^{\alpha_\ell}}, & \text{if } \sigma_{\mathrm{Net},\ell} < \text{T}\,,
\end{cases}
&\ell < 30, \\
\displaystyle \frac{1}{\sqrt{2\pi\,\sigma_{\mathrm{total},\ell}^2}} \exp\!\left[-\frac{\bigl(C_\ell^{\mathrm{pred}}-C_\ell^{\mathrm{theory}}\bigr)^2}{2\sigma_{\mathrm{total},\ell}^2}\right],   &\ell\ge 30\,,
\end{cases}
\end{align}}
where, $\alpha_\ell = \frac{2\ell + 1}{2}$, $\theta_\ell = \frac{2C_\ell^{\text{theory}}}{2\ell + 1}$, 
$T = 0.2\sqrt{\theta_\ell^2 \alpha_\ell}$ and
$\sigma_{\text{total},\ell}^2 = \theta_\ell^2 \alpha_\ell + \sigma_{\text{Net},\ell}^2$.
For low multipoles, $\ell$ < 30, we perform an adaptive integration over a grid of $C_\ell$ values in the range 
defined by $C_\ell^{\tt pred} \pm 5\theta_\ell^2 \alpha_\ell$. At every grid point, we calculate the 
logarithm of both Gamma probability density function which represents the cosmic variance and 
Gaussian probability density function which takes care of the network error. We keep a threshold for network 
error, i.e., if network error is less than 20$\%$ the cosmic variance at that mulitpole, then network error will 
not be considered. This is in order to prevent any numerical instabilities due to very small network error.  
For multipoles $\ell\, \ge\, 30$, we adopt a chi-square likelihood with variance as the quadrature sum of 
cosmic variance and the network error. 
The likelihood evaluated following this approach is displayed in Fig.~\ref{like} for 
$r$-values 0.0121, 0.00445 and 0.000645. The maximum likelihood values of $r$ and $A_L$ is very close to the 
true value signifying our networks predicitions along with the error estimates is a valuable tool in 
minimizing the foreground and noise in the observed CMB B-mode maps.

\section{Conclusions \& Discussions}
\label{Conclusion}
By incorporating Bayesian machine learning techniques in conjunction 
with U-Net and ResNet architectures, our network ${\tt PUREPath-B}$, offers a powerful framework 
to minimize the foreground contaminations in CMB B-mode maps.  
The uncertainty estimates provided by our network by leveraging its inherent capability 
to quantify the uncertainty in its predictions can be directly incorporated 
into subsequent cosmological analyses like parameter estimation etc. 

Our network is trained using 3000 sets of simulated foreground contaminated CMB maps 
at 8 LiteBIRD frequency channels. We use thermal dust, synchrotron  
and AME as the major sources of foreground contaminations. 
During training, the prior distributions over model 
parameters in our network acts as regularization, especially when the training data is 
limited or noisy.  By utilizing Convolution2DFlipout layers and VI 
techniques, our network efficiently approximate complex posterior distributions of the 
model parameters.
During inference, we implement the testing dataset to evaluate the preformance of our network. We 
estimate predictive mean cleaned maps and its power spectrum correspoding to different values of 
$r$ and $A_{\tt Lens}$. 
The predictive mean maps matches quite well with the  input CMB B-mode map used in those simulations. 
The Error map quantifies our model's uncertainty  on the predicted cleaned CMB 
map. This is particularly useful in contexts like using our 
cleaned map for cosmological 
parameter estimation and other cosmological analyses where 
understanding the uncertainty in the predicted map is as crucial as the 
predictions themselves. We provided a method to include the estimated network error to the 
paramater estimation pipeline. 

Overall, the proposed approach provides a comprehensive solution for 
minimizing foreground and noise contaminations in  CMB B-mode maps, enabling more accurate measurements 
of cosmological 
parameters and other cosmological analyses. 

\section{Acknowledgments}
 
We  use publicly available HEALPix~\cite{Gorski:2004by} package  
(http://healpix.sourceforge.net) for the analysis of this work.  The network 
we have developed is based on 
the libraries provided by Tensorflow and Tensorflow Probability.

\bibliography{ms.bib}{}
\bibliographystyle{aasjournal}

\end{document}